\documentclass[preprint,12pt,authoryear]{elsarticle}

\usepackage{amssymb}
\usepackage{amsmath}

\journal{Solar Energy}

\begin{document}

\begin{frontmatter}



\title{Adapting Quantile Mapping to Bias Correct Solar Radiation Data} 


\author[csm,nrel]{Maggie D. Bailey}
\author[csm]{Douglas W. Nychka}
\author[nrel]{Manajit Sengupta}
\author[csm]{Soutir Bandyopadhyay} 

\affiliation[csm]{organization={Colorado School of Mines},
            addressline={1500 Illinois Street}, 
            city={Golden},
            postcode={80401}, 
            state={Colorado},
            country={USA}}

\affiliation[nrel]{organization={National Renewable Energy Lab},
            addressline={15013 Denver West Parkway}, 
            city={Golden},
            postcode={80401}, 
            state={Colorado},
            country={USA}}

\begin{abstract}
Bias correction is a common pre-processing step applied to climate model data before it is used for further analysis. This article introduces an efficient adaptation of a well-established bias-correction method, quantile mapping, for global horizontal irradiance (GHI) that ensures corrected data is physically plausible through incorporating measurements of clearsky GHI. The proposed quantile mapping method is fit on reanalysis data to first bias correct for regional climate models (RCMs) and is tested on RCMs forced by general circulation models (GCMs) to understand existing biases directly from GCMs. Additionally, we adapt a functional analysis of variance methodology that analyzes sources of remaining biases after implementing the proposed quantile mapping method and consider biases by climate region. This analysis is applied to four sets of climate model output from NA-CORDEX and compared against data from the National Solar Radiation Database produced by the National Renewable Energy Lab.
\end{abstract}

\begin{graphicalabstract}
\includegraphics[width=\textwidth]{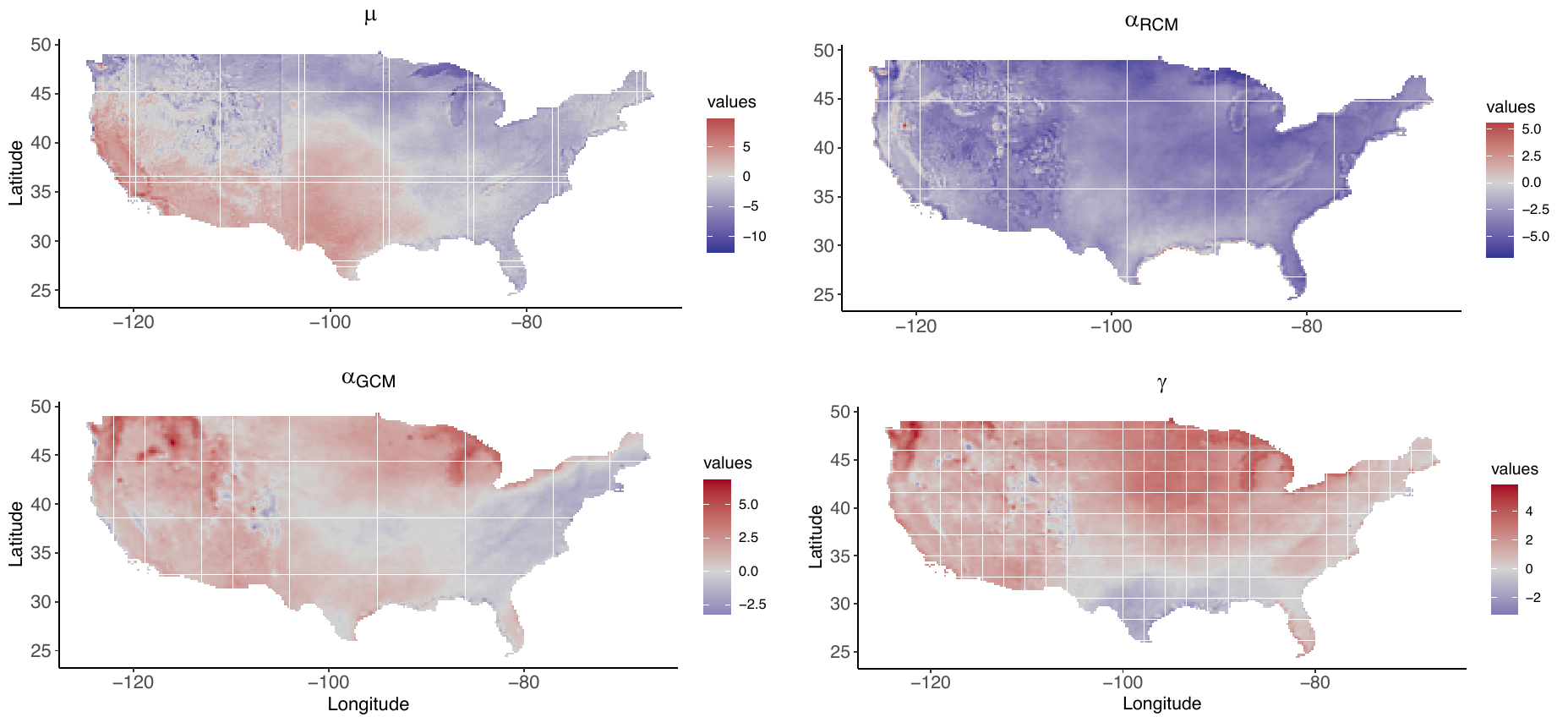}
\end{graphicalabstract}

\begin{highlights}
\item Adaptation of quantile mapping specific to GHI that considers clearsky GHI does well to preserve physically plausible GHI values 
\item Geographic specific biases appear to be minimal for RCMs considered
\item RCMs seem to contribute most to biases and should be corrected for before downstream use
\end{highlights}


\end{frontmatter}



\section{Introduction}
\label{sec:bc_intro}

Quantile mapping is a well-established method that has been applied to a variety of climate variables, including temperature, precipitation, and solar radiation \citep{frank2018bias, kakimoto2022quantile, bennett2014performance}. Additionally, numerous extensions and adaptations have been proposed for improved applications to specific variables and to address common criticisms of the method, including not accounting for spatial or temporal aspects of climate data \citep{maraun2016bias}. For example, quantile mapping can be extended to adapt to seasonal changes in temperature and precipitation after downscaling has been implemented using local regression \citep{washington2023modeling}. This builds upon earlier research which adapted quantile mapping methods to monthly and moving window correction \citep{thrasher2012bias, wood2002long}. Quantile mapping has also been extended to bias correct multiple variables from a single climate model to preserve dependence between variables \citep{cannon2018multivariate}. The ratio between corrected and raw modeled data, defined as \textit{change rates} in this study, may also be calculated through quantile mapping and applied to future years to correct model data \citep{yuan2016bias}. Within the context of solar radiation, extensions
of quantile mapping have shown to improve on other bias correction methods. For example, considering physical upper limits of radiation, here empirically estimated using 25-day running maximum values from multi-year maximum datasets, and correcting at the daily level over monthly has shown to be more effective in non-parametric quantile mapping \citep{lange2018bias}.

Similar to these studies, this article proposes an extension to quantile mapping to improve bias correction for gridded solar radiation data. In addition to adjusting distributions on a monthly scale, we introduce a pre-processing step to solar radiation data to ensure the final bias corrected data will not exceed clearsky GHI, a quantity assuming clear sky conditions that represents maximum possible GHI at the surface calculated from satellite measurements. We address spatial variations in solar radiation bias sourced from climate model data through correcting on a pixel by pixel level while considering the distribution of solar radiation at a pixel's nearest neighbors. A primary goal of this analysis is to understand how well GCMs predict solar radiation after being dynamically downscaled using RCMs, while correcting for existing biases from RCMs. We assess where and when remaining biases typically occur in climate model data and whether these biases are effects from RCMs, GCMs, or an interaction between both. 

The rest of the article is organized as follows: in Section~\ref{sec:bc_data} we describe the data used in this study. Section~\ref{sec:bc_methods} introduces the adaptation for quantile mapping and results for this method are shown in Section~\ref{sec:bc_results}. Finally, we conclude the aticle and propose future research directions in Section~\ref{sec:bc_conclusions}.

\section{Data}
\label{sec:bc_data}

\subsection{National Solar Radiation Database}
\label{sec:bc_data_nsrdb}

The NSRDB provides satellite modeled solar radiation data for the three most common types of solar radiation variables (GHI, DNI, and DHI measured in $Wm^{-2}$) at hourly and half-hourly time scales for years 1998-2022. These variables are calculated using the Physical Solar Model which takes as input cloud properties derived from satellite measurements
\citep{sengupta2018national}. For the purposes of this article, the dataset is averaged to a grid resolution of 20 km and a daily time scale to be comparable with climate model output at a similar resolution.

A focus of this method is to provide bias corrected GHI that is physically plausible. To do this, we calculate the clearsky index ($k_c$), the ratio between GHI and clearsky GHI, for daily average bias corrected GHI. Clearsky GHI is part of the NSRDB data product and is a modeled quantity estimating the maximum possible GHI at the surface of the earth. This value can change depending on atmospheric conditions. Therefore, we generated a clearsky dataset that takes the average of clearsky values across 23 years of data from the NSRDB (1998-2020). At the time this was calculated, only data up through 2020 was available. Since this is an average of modeled quantities, clearsky index  values calculated after quantile mapping that exceed 1 are possible but should be sparse and be commensurate with the same number of values over 1 as seen in the NSRDB. The average clearsky GHI calculated here is referred to as the empirically derived clearsky dataset throughout this chapter. 

\subsection{Climate Model Database}
\label{sec:bc_data_nacordex}

We pre-process climate model data by regridding it to the same grid as the NSRDB using thin plate splines. The uncertainty in this step has been studied previously and shown to have little effect on downstream modeling \citep{bailey2023regridding}. All data from this point forward is on a common 20 km grid and at a daily time scale. Climate model data comes from the NA-CORDEX data archive, containing RCM output forced by various GCMs \citep{mcginnis2021building}. The archive includes RCMs forced by the reanalysis ERA-Interim generally covering 1979-2014 and GCM driven runs that extend from 1950-2100. The GCMs used as boundary inputs for the RCMs are from the Couple Model Intercomparison Project Phase 5 (CMIP5). While there are newer GCMs available through Phase 6, the availability of updated RCMs forced by newer GCMs is limited. NA-CORDEX provides a wealth of RCMs with various forcings which can be used to analyze existing biases in both RCMs and GCMs from CMIP5. The ERA-Interim driven runs are used in this section to focus bias-correction to correcting RCM biases first and then analyzing existing biases from GCMs. See Table~\ref{tab:nacordex_models} for key model characteristics of the RCMs chosen for analysis in this chapter. Throughout the rest of this article, specific climate model datasets will be referred to using the acronyms for the GCM and RCM, for example the RegCM4 RCM which has boundary input provided by the HadGEM2-ES GCM will be referred to as the HadGEM2-ES.RegCM4.

\begin{figure}[t]
    \centering
    \includegraphics[width = \textwidth]{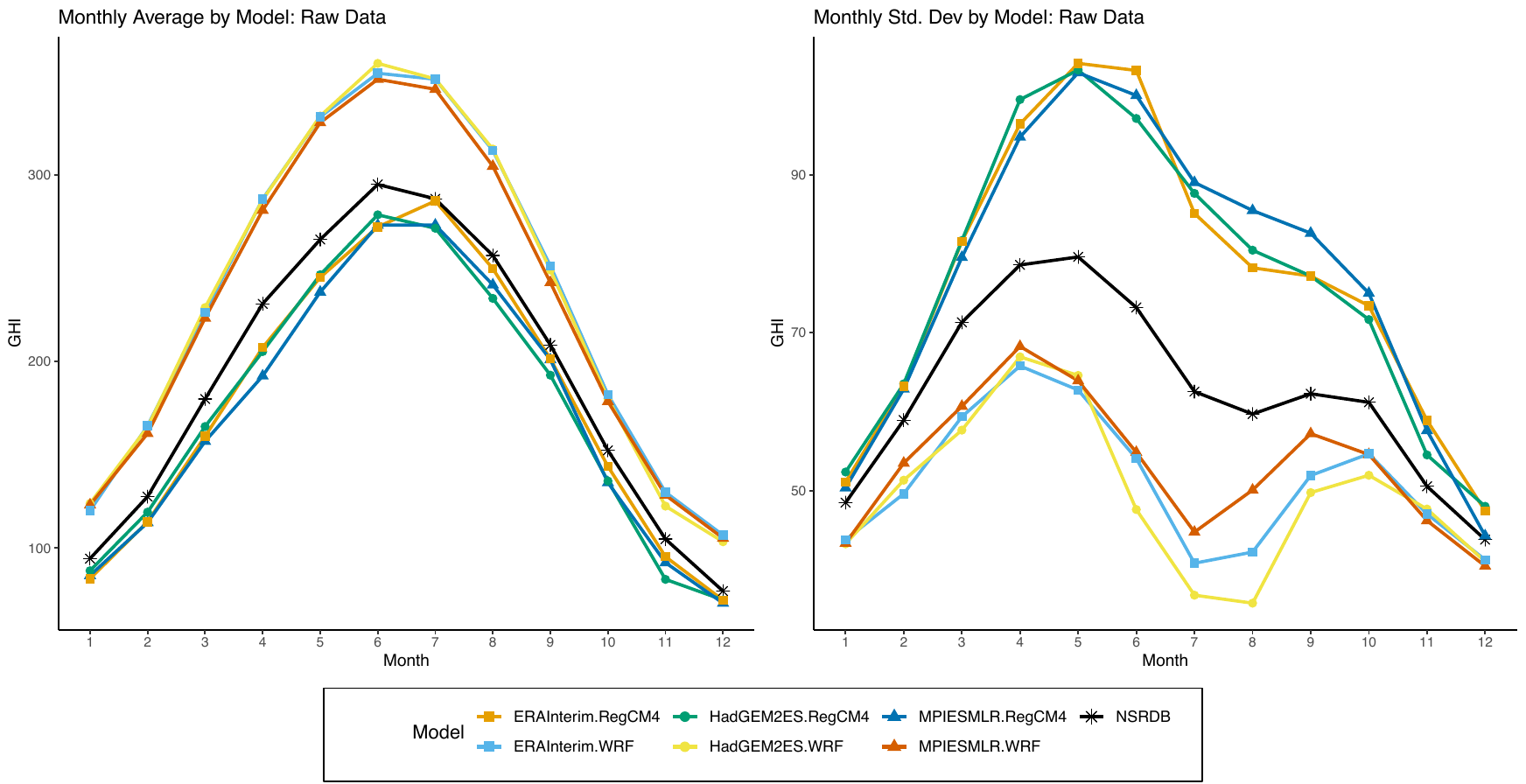}
    \caption{Monthly average data for NSRDB (2011-2015), GCM driven data (2011-2015), and ERA-Interim data (1998-2010).}
    \label{fig:monthly_avg_sd_2011_2015}
\end{figure}

Monthly mean and standard deviations for uncorrected NA-CORDEX data compared to NSRDB data are presented in Figure~\ref{fig:monthly_avg_sd_2011_2015}. Here we see that model data including the WRF RCM tend to have a positive bias for the mean but a negative bias for the standard deviations. Surprisingly, the opposite is true for data from the RegCM4 RCM. These biases are consistent regardless of which GCM is used as boundary input for the RCM.

\begin{table}[t]
	\caption{\label{tab:nacordex_models} Model data used with corresponding attributes for the bias-correction analysis in this chapter.}
\begin{center}
\begin{tabular}{||c c c c c c||} 
 \hline
 Model & Driver & RCP & Years  & Grid Resolution & Calendar\\ [0.5ex] 
 \hline\hline
 NSRDB & NA & NA & 1998-2022 & 4 km  & Gregorian  \\
 \hline
  NSRDB (aggregated) & NA & NA & 1998-2022 & 20 km & Gregorian  \\
 \hline 
 RegCM4 & ERA-Interim & NA & 1979-2014 & 25 km & Gregorian\\
 \hline
 RegCM4 & HadGEM2-ES & 8.5 & 2006-2099 & 25 km & 360 day\\
 \hline
 RegCM4 & MPI-ESM-LR & 8.5 &  2006-2100 & 25 km & Gregorian \\
 \hline
 WRF & ERA-Interim & NA & 1980-2010 &  25 km & Gregorian\\
 \hline
 WRF & HadGEM2-ES & 8.5 & 2006-2099 & 25 km & Gregorian \\
 \hline
 WRF & MPI-ESM-LR & 8.5 & 2006-2100 & 25 km & 365 day\\
 \hline
\end{tabular}
\end{center}
\end{table}

\section{Methods}
\label{sec:bc_methods}

\begin{figure}
    \centering
    \includegraphics[width=0.5\textwidth]{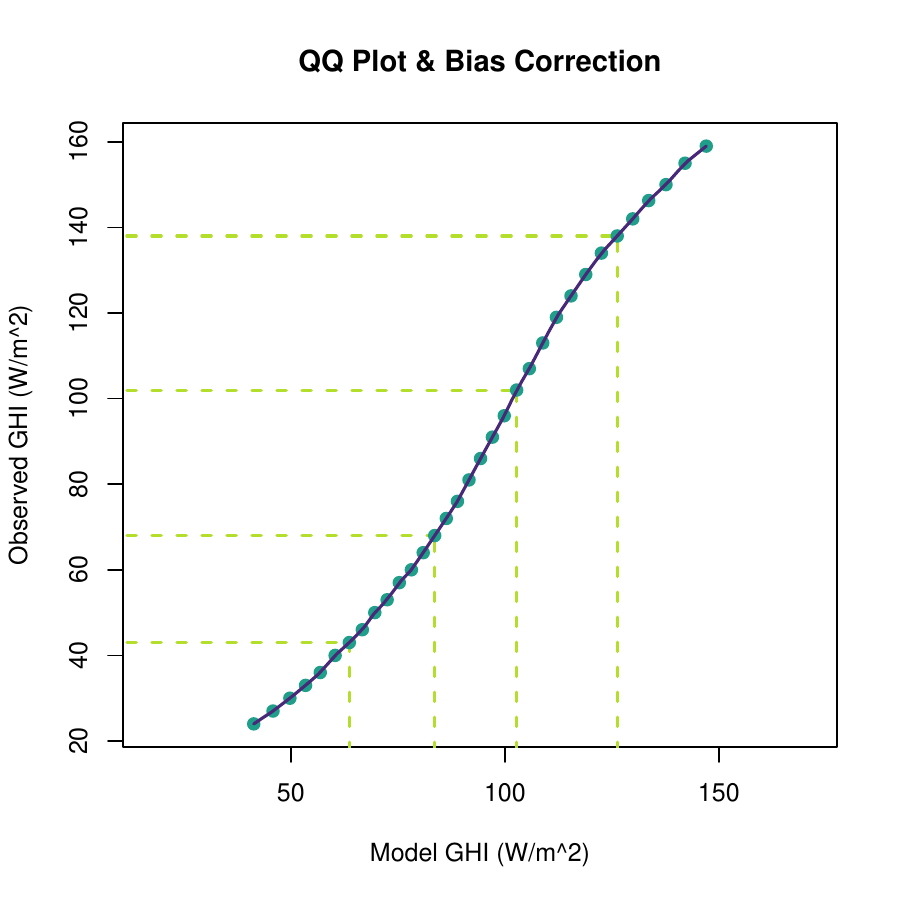}
    \caption{Illustration of how quantile-quantile mapping adjusts the distribution of modeled GHI according to the quantiles of the observed distribution}
    \label{fig:qqmap_example}
\end{figure}

Quantile mapping is a widely used bias correction method in climate science based on an empirical transformation from one distribution to another. A visual representation of how the model distribution is corrected against the observed distribution is shown in Figure~1993\ref{fig:qqmap_example}. At the most basic level, given the empirical cumulative distribution function (CDF) of an observed dataset, $\hat{F}_{obs}$, and empirical CDF for model data, $\hat{F}_{mod}$, the quantile map is 
\begin{equation}
    \hat{F}^{-1}_{obs}(\hat{F}_{mod}(x)) = T(x).
    \label{eq:quantile_map}
\end{equation}

The model results are then bias corrected by applying $T(\cdot)$ to the model output. The function $T(\cdot)$ is sometimes known as a \textit{transfer function} \citep{maraun2016bias}. Note that in general, $T(\cdot)$ will map the $\alpha$ quantile for the model output to the $\alpha$ quantile of the observed data. The deviation of $T(\cdot)$ from an identity map results in the bias correction. If the CDF has an assumed parametric distribution, such as the normal distribution, the method is called parametric quantile mapping. Non-parametric quantile mapping assumes no distribution and the distribution is instead empirically estimated through a selected set of quantiles. The exact quantiles are chosen based on the application and in this application, the CDF is empirically estimated using a set of equally spaced quantiles:
\begin{equation*}
    [0.01, 0.02, 0.03, \ldots, 0.97, 0.98, 0.99].
\end{equation*}
Thus, the complete empirical $T(\cdot)$ is reduced to estimating its value at these 99 quantiles. Given a discrete set of quantiles, $T(\cdot)$ is approximated by linear interpolation between each quantile. There are other types of interpolation available for this step, such as cubic smoothing splines, however we found linear interpolation corrects accurately for intermediate values, particularly with using a large number of quantiles, avoids extrapolation issues at boundary quantiles, such as below $0.01$ or above $0.99$, and also maintains monotinicity of $T(\cdot)$. Additionally, linear interpolation is quick to estimate and predict and therefore efficient for a larger gridded datasets, such as the climate model data in this application. For GHI values that fall above the $0.99$ or below the $0.01$ quantiles, we linearly extrapolate in either direction based on interpolating the two lowest and two highest unequal quantile values from the quantile quantile plot. Note this assumes a linear transformation, a location and scaling, beyond these quantile ranges.

For this solar application we require an extension of quantile mapping to ensure that the resulting bias corrected data are physically plausible. The clearsky index is first calculated for daily average GHI data for both climate model and observed data by dividing by the daily average clearsky maximum GHI ($CS^{i,d}$) for location $i$ and day $d$:
\begin{equation*}
    k_c^{i,d} = \frac{GHI^{i,d}}{CS^{i,d}} \label{eq:kc_ratio}
\end{equation*}
where the resulting values $k_c^{i,d}\in [0,1]$. If the $k_c$ value calculated from a climate model is greater than 1, it is capped at 1. This ensures that climate model data to be corrected remains below the physically possible clearsky maximum. Because clearsky GHI from the NSRDB is a modeled quantity obtained by assuming clearsky conditions on a given day and for a particular location, it may change for the same location across years depending on atmospheric conditions. Therefore, resulting bias-corrected values for testing years may still fall above the clearsky maximum but should still be reasonably valued. No values should fall below zero. To control for changes in clearsky GHI from year to year, we use the empirically derived clearsky GHI dataset described in Section~\ref{sec:bc_data_nsrdb}. The logit transformation is applied to the clearsky index calculated for observed and climate model data to transform values to be in $[-\infty, \infty]$:
\begin{equation}
    t^{i,d}=logit(k_c^{i,d}) =log\left(\frac{k_c^{i,d}}{1-k_c^{i,d}}\right). 
\end{equation}
Let $\boldsymbol{t}^{i}_{mod}$ be the vector of transformed clearsky values for the climate model data for location $i$ and days $d=1, \dots, D$, where $D$ is the number of days for a single month. Let $\boldsymbol{t}^{i}_{obs}$ be the same for the observed data. $\hat{F}_{obs}$ is estimated for  $\boldsymbol{t}^{i}_{obs}$ and $\hat{F}_{mod}$ for $\boldsymbol{t}^{i}_{mod}$ for observed years, or in this analysis training data.  Equation~\ref{eq:quantile_map} is applied to $\boldsymbol{t}^{i}_{mod}$ for unobserved years or testing data. As mentioned earlier, values that fall below or above the 0.01 and 0.99 quantiles are linearly extrapolated beyond the 0.01 and 0.99 quantile. While this increases or decreases the input climate model value beyond the observed 0.99 and 0.1 quantiles, quantile mapped values are reverted back to be in $[0,1]$ using the logistic and then multiplied by the clearsky GHI value for location $i$ and day $d$. This results in physically plausible, bias corrected GHI daily averages given a daily average GHI value from a climate model.


To ensure smooth transitions across pixels, we implement a tiling approach where quantile mapping is applied by month to each individual pixel and its nearest neighbors. That is, for a given location and month, daily GHI data for the eight nearest neighbors, resulting in nine pixels each with 30 days of data, is used in estimating the CDFs $\hat{F}_{mod}$ and $\hat{F}_{obs}$. We found that including a higher number of nearest neighbors does not significantly improve or change the empirical CDF.

\subsection{Analysis Design}
\label{sec:bc_analysis_design}

We implement the adapted quantile mapping model described in the previous section for daily average GHI for the six climate datasets outlined in \ref{tab:nacordex_models} in two ways. First, as a form of cross-validation, we fit the quantile mapping method on ERA-Interim driven RCMs for 2001-2005 and correct out of sample ERA-Interim data for 2006-2010. Bias corrected ERA-Interim driven RCM data is then compared to observed NSRDB data from the same years. This is done to asses the quantile mapping method for ERA-Interim driven RCMs for out of sample years. ERA-Interim is a global atmospheric reanalysis dataset and can be thought of as a ``perfect" GCM as the data represents observed weather patterns. By performing this type of cross-validation, we will be able to assess the method's ability to correct specifically for RCM biases in the out of sample years. By considering individual pixels and their nearest neighbors to estimate the empirical CDF, as described in the previous section, this design provides approximately $30 \text{ days}\times 5 \text{ years} \times 9 \text{ pixels}  = 1,350$ daily averages to estimate an empirical CDF for each pixel in CONUS.

Second, RCMs that use GCMs as boundary input are corrected using the adapted quantile mapping method. For this analysis, the reanalysis driven RCMs are first subsetted to 1998-2010, or all overlapping years from the ERA-Interim driven RCMs, for fitting the quantile map. This fit is then used to correct four datasets subsetted to years 2011-2015, based on a $2\times 2$ factorial design with the factors being choice of GCM and choice of RCM: RegCM4 and WRF each forced by HadGEM2-ES and MPI-ESM-LR. This allows for an intercomparison analysis on the effectiveness of bias-correcting directly for RCM biases while considering remaining GCM biases. The design also ensures out of sample prediction on years 2011-2015 for the GCM models compared to observed data for the same years.

Quantile mapped climate model data are first analyzed through monthly means and standard deviations for all CONUS. Second, we consider annual average GHI by year, model, geography, and by Bukovksy climate regions \citep{bukovsky2012masks}. Bukovsky climate regions were established for analyzing results from the North American Regional Climate Change Assessment Program \citep{mearns2009regional} as a logical division of North America into terrestrial eco-regions without resorting to a pixel by pixel comparison. They closely follow regions from the National Ecological Observatory Network and are defined on a common $1/2 \times 1/2$ degree grid. The climate region for each NSRDB pixel is determined by the closest Bukovsky region to the NSRDB pixel in terms of Euclidean distance. There are 20 distinct Bukovsky regions for the NSRDB in CONUS used for this analysis, shown in Figure~\ref{fig:bukovsky_regions_conus}.

\begin{table}[t]
    \centering
    \caption{The $2\times 2$ factorial analysis design. Here $\beta_{ij}$ is the bias field (i.e. the collection of biases by NSRDB pixel) resulting from that particular RCM/GCM combination for $i,j = 1,2$.}
    \begin{tabular}{ c | c  c }
            & HadGEM-ES2 & MPI-ESM-LR \\
        \hline
        WRF & $\beta_{11}$ & $\beta_{12}$  \\
        RegCM4 & $\beta_{21}$ & $\beta_{22}$ \\
    \end{tabular}
    \label{tab:factorial_design}
\end{table}

Finally, we adapt a functional analysis of variance (FANOVA) method to assess remaining sources of bias in GCM driven runs \citep{sain2011functional}. A FANOVA approach allows inference on sources of remaining biases after quantile mapping is implemented. Such analyses could be used as a tool for climate model developers to understand where remaining biases exist or help identify why biases persist after correction. The FANOVA method can be briefly summarized as an analysis of a 2$\times 2$ factorial experiment ``pixel by pixel," as defined in Table~\ref{tab:factorial_design}. Note that the layout specifies GCMs as the columns and RCMs are the rows. Each $\beta_{i,j}$ for $i,j = {1,2}$, is a particular field of biases after implementing the bias correction model. We define a baseline bias, or overall bias, as the mean across all $\beta_{i,j}$:
\begin{equation*}
    \mu = \frac{\beta_{11} + \beta_{21} + \beta_{12} + \beta_{22}}{4}.
\end{equation*}
To estimate the main effects and the interaction from the RCMs and GCMs, we define
\begin{align*}
    \alpha_{GCM} &= \frac{\beta_{12} - \beta_{11} + \beta_{22} - \beta_{21}}{4}\\
    \alpha_{RCM} &= \frac{\beta_{21} + \beta_{22} - \beta_{11} - \beta_{12}}{4}\\
    \gamma &= \frac{\beta_{11} - \beta_{12} + \beta_{22} - \beta_{21}}{4}
\end{align*}
where $\alpha_{RCM}$ and $\alpha_{GCM}$ will describe the main effects on remaining biases from RCMs and GCMs, respectively. The final term, $\gamma$, is an interaction term which will describe the main effects from the interaction between RCMs or GCMs. Put another way, $\gamma$ describes whether particular RCM and GCM combinations produce significant bias compared to the RCMs and GCMs alone.

The FANOVA analysis is applied to bias corrected GCM and RCM combinations listed in Table~\ref{tab:nacordex_models} for 2011-2015. The bias fields are averaged across the four years and the $2\times 2$ factorial analysis is then performed on the average bias from these cases. In this way, we can gain a broader understanding of where model biases persist over several years since we don't expect each year to match exactly with observed data.


\section{Results}
\label{sec:bc_results}

\subsection{Monthly mean and standard deviation}
\label{sec:bc_results_mon_avg_sd}

\begin{figure}
    \centering
    \includegraphics[width=\textwidth]{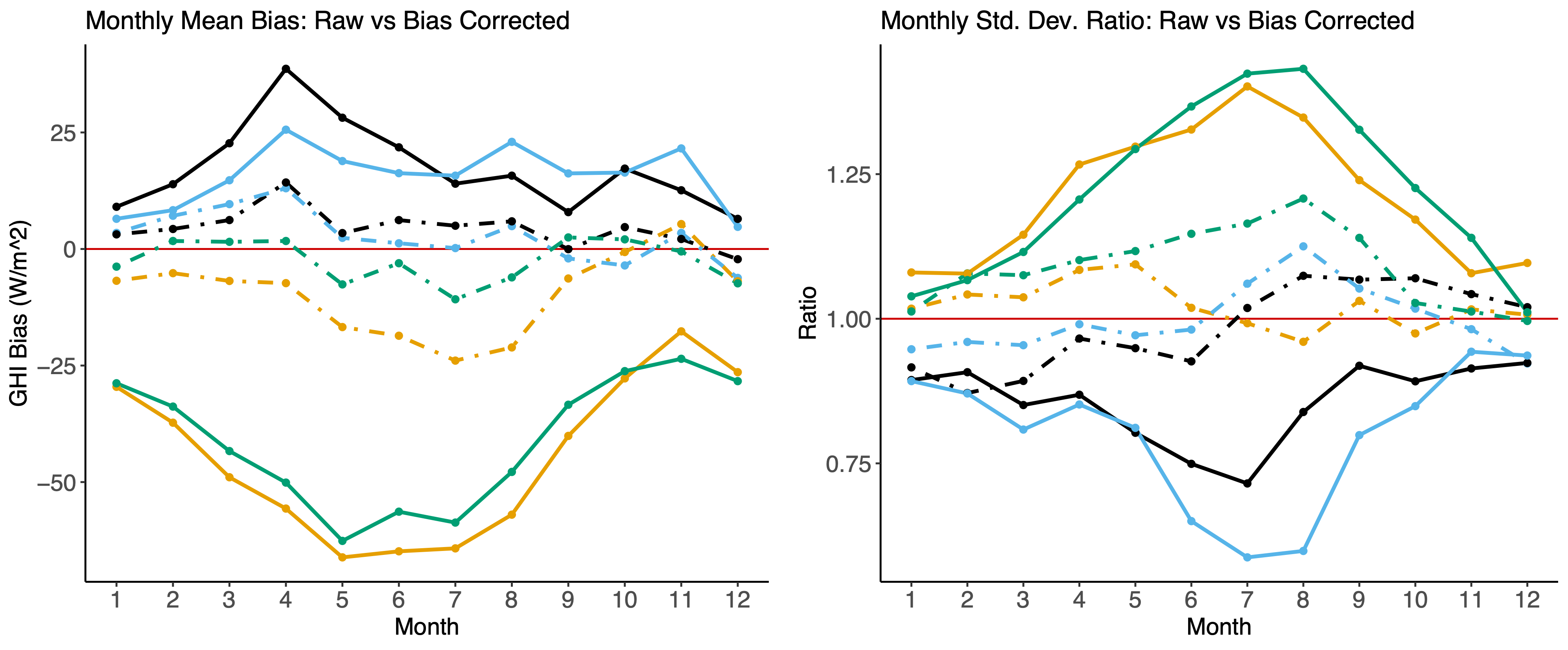}
    \\[0.5cm]
    \includegraphics[width=\textwidth]{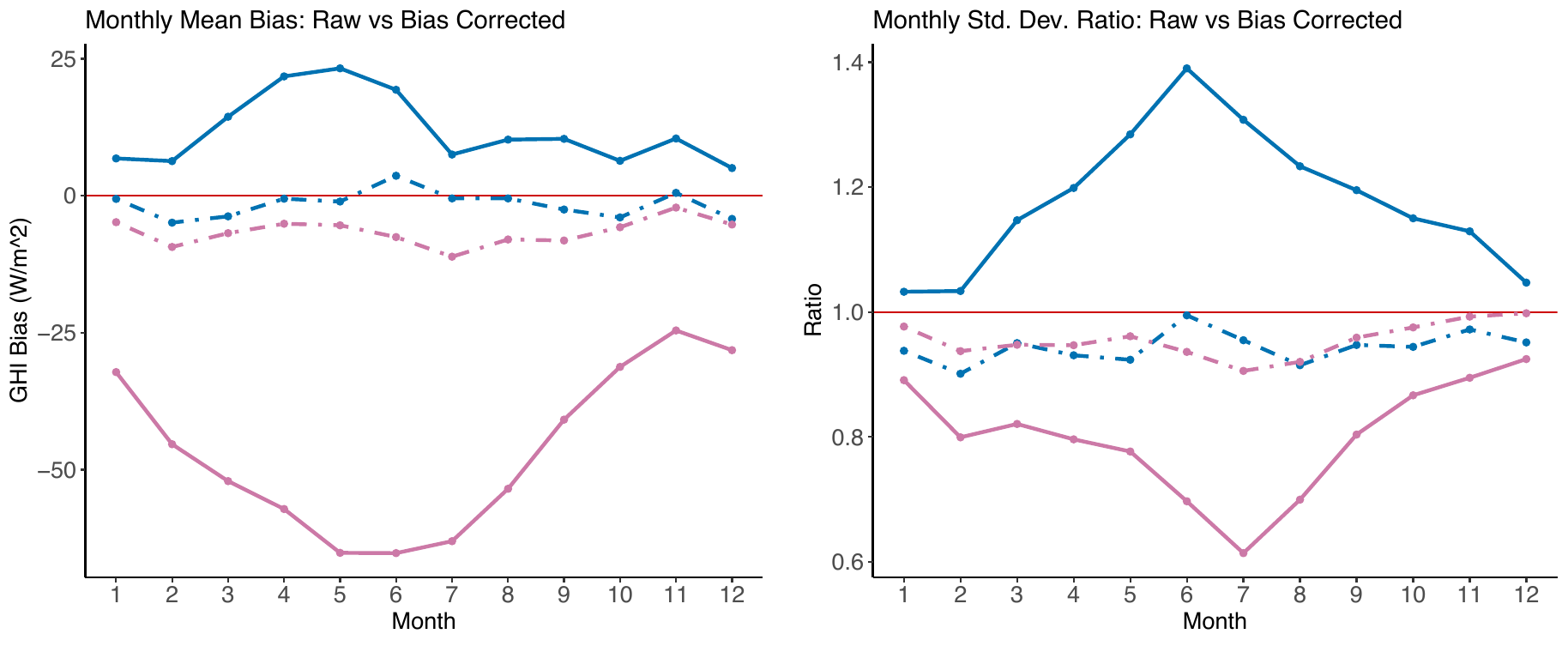}
    \caption{Top: bias between monthly means (left) and the ratio of monthly standard deviations (right) between NSRDB and RCM data for GHI for 2011-2015. Raw data (solid lines) and bias corrected (dot-dash lines) are plotted together for four models: HadGEM2-ES.RegCM4 (black), MPI-ESM-LR.RegCM4 (light blue), HadGEM2-ES.WRF (orange), and MPI-ESM-LR.WRF (green). Bottom: raw (solid lines) and bias corrected (dot-dash lines) monthly mean bias (left) and standard deviation ratios (right) for ERA-Interim.WRF (pink) and ERA-Interim.RegCM4 (dark blue) for out of sample prediction in years 2006-2010.}
    \label{fig:monthly_mean_sd_bias}
\end{figure}

Figure~\ref{fig:monthly_mean_sd_bias} shows biases before and after quantile mapping for monthly mean and standard deviations across the four models considered for out of sample, cross-validation comparisons. The left column shows biases in the monthly mean and the right column shows biases in monthly standard deviation. The lines represent the bias between the monthly averages calculated as the climate model GHI data subtracted from the observed GHI data. Any negative values indicate the modeled data overpredicts the value plotted while positive data indicates the opposite. 

Figure~\ref{fig:monthly_mean_sd_bias} shows that uncorrected climate model biases largely result from the choice of RCM. For example, uncorrected climate model data from the WRF RCM for both GCM forcings tends to overpredict monthly means compared to observed data, especially in summer months where the bias reaches over 50 $W/m^2$. On the other hand, uncorrected data from the RegCM4 RCM for both GCM forcings tends to underpredict monthly average GHI values, but more closely aligns with observed data with a maximum bias of nearly $39\; W/m^2$ from HadGEM2-ES.RegCM4. The proposed quantile mapping method does well in correcting these biases, where the dashed lines show the monthly mean bias for bias-corrected data. Bias corrected data remains within $[-25\;W/m^2, 14.3\;W/m^2]$ compared to $[-66.2\;W/m^2,  38.7\;W/m^2]$, reducing the bias by about 63\% overall.

The top right plot in Figure~\ref{fig:monthly_mean_sd_bias} shows the monthly standard deviation by choice of GCM and RCM combination. Here we see similar trends in the uncorrected versus corrected bias for standard deviation in that the choice of RCM drives resulting biases.  The RegCM4 RCM tends to overpredict the monthly variability while WRF tends to underpredict. This is true regardless of the GCM forcing. Again, we see that the proposed quantile mapping method corrects for larger biases, especially during summer months, reducing bias from about $\pm 25 \; W/m^2$ to within $[-12.4, 7.7]$ $W/m^2$. 

For ERA-Interim only bias correction, seen in the bottom row of Figure~\ref{fig:monthly_mean_sd_bias}, we see that the method corrects well for RegCM4 monthly means but there are remaining biases in the monthly means for WRF after quantile mapping. However, there is a significant reduction the bias for WRF where initial, uncorrected monthly mean bias reached over 50 $W/m^2$. For monthly standard deviations, there is again a significant reduction in bias. However, we see that there are some remaining biases in variability through the year. In particular, the bias-corrected data underpredicts monthly variability compared to observed data, though there are no distinct patterns for particular months unlike the uncorrected data.

\subsection{Annual average GHI}
\label{sec:bc_results_ann_avg_ghi}

The percent bias remaining in annual average GHI ($kWh/m^2/day$) for entire CONUS for 2011 is shown in Figure~\ref{fig:ann_avg_ghi_bias_2011}. Additional years are shown in \ref{appendix:bc_additional_figures}. Because the method bias corrects for biases in the RCMs directly, where each RCM is forced by the reanalysis data, ERA-Interim, we can consider remaining differences as artifacts of systematic biases in the GCMs. This will be further explored from FANOVA results in Section~\ref{sec:bc_results_fanova}. Generally, all models tend to underpredict annual average GHI in northern CONUS and underpredict in southern CONUS. The strength of this trend and smaller scale geographic trends seems to be driven by the RCM. This makes sense as RCMs are developed to model finer-scale climate trends within a particular domain. Some areas of CONUS still have biases in areas of varied terrain, which can be seen in western CONUS across all bias fields. 

\begin{figure}
    \centering
    \includegraphics[width=\textwidth]{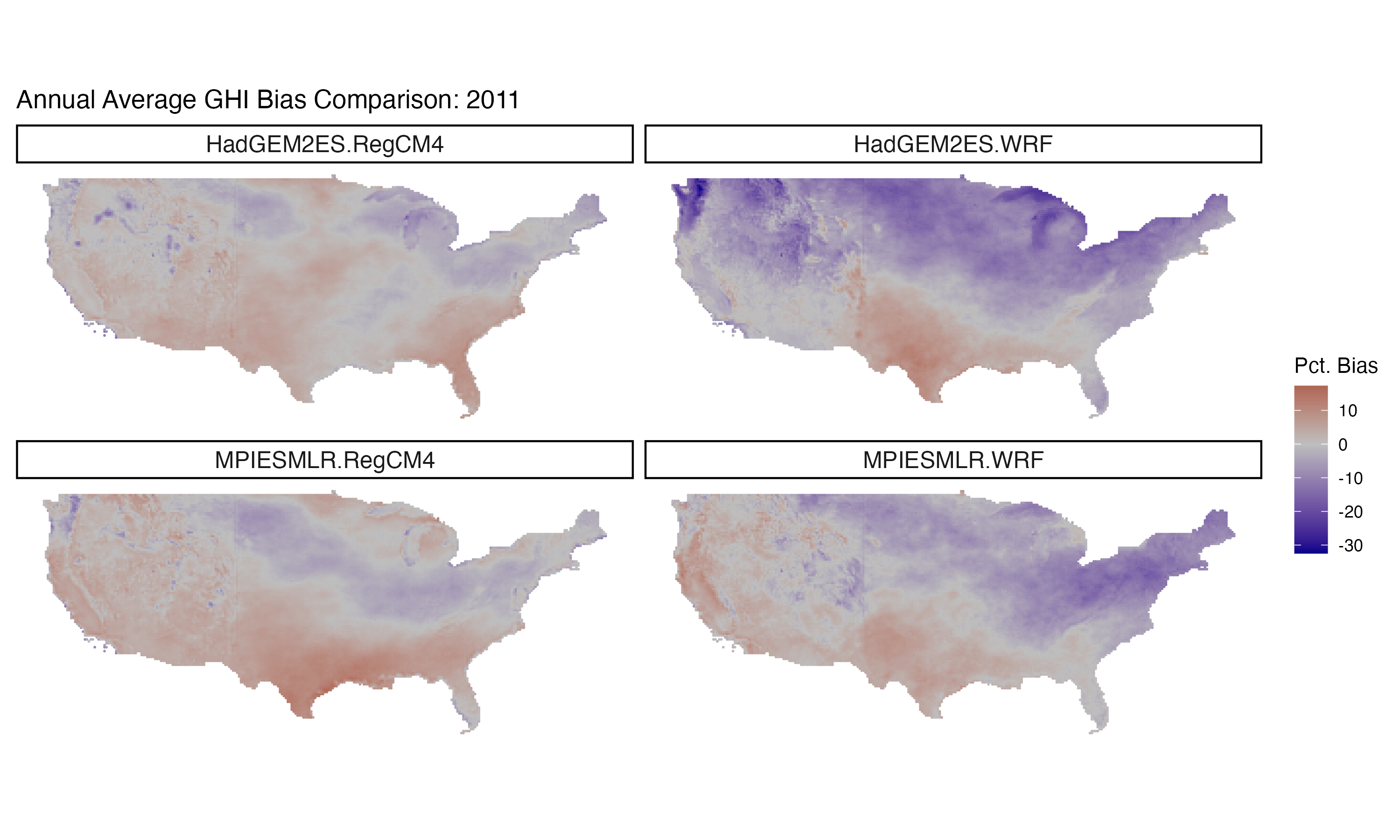}
    \includegraphics[width=\textwidth]{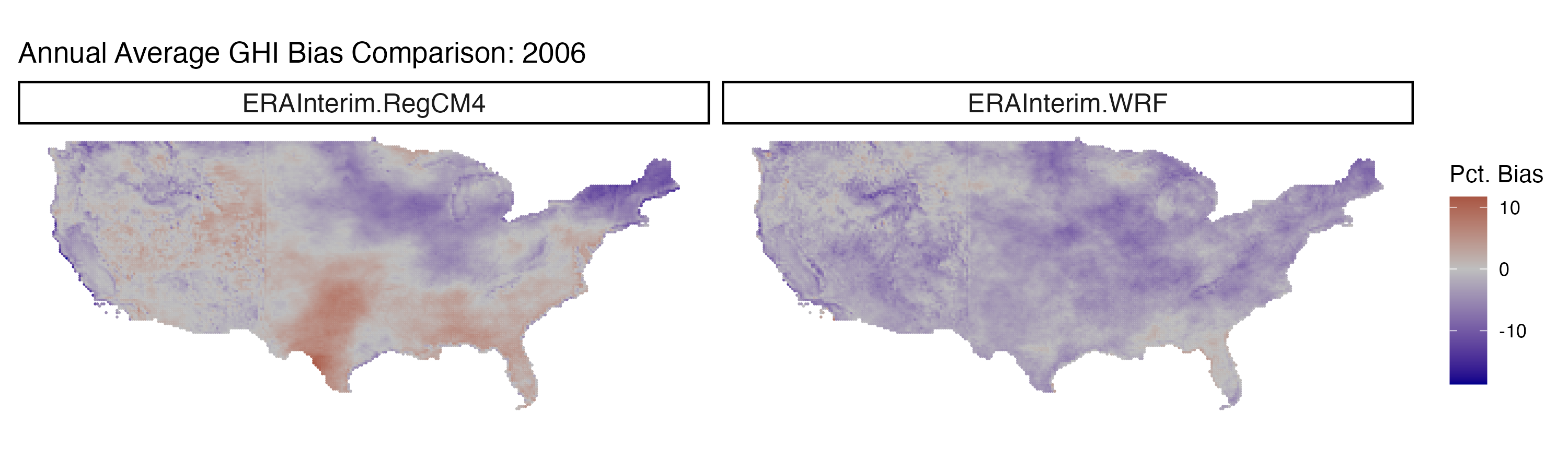}
    \caption{Percent bias across CONUS in annual average GHI for all GCM models in 2011 (top figure) and out of sample ERA-Interim in 2006 (bottom figure) after implementing proposed quantile mapping.}
    \label{fig:ann_avg_ghi_bias_2011}
\end{figure}

For ERA-Interim only driven runs, we see that bias corrected data for RegCM4 in 2006 is centered at zero but there remains over- and underestimation of annual average GHI in similar areas seen for the GCMs. Similar to monthly mean bias, we again see that bias corrected WRF data still over predicts in many areas of CONUS, though there is not a distinct geographic pattern of where this tends to occur except for areas in in California near the Sierra Nevadas, for example.

\begin{figure}
    \centering
    \includegraphics[width=0.8\textwidth]{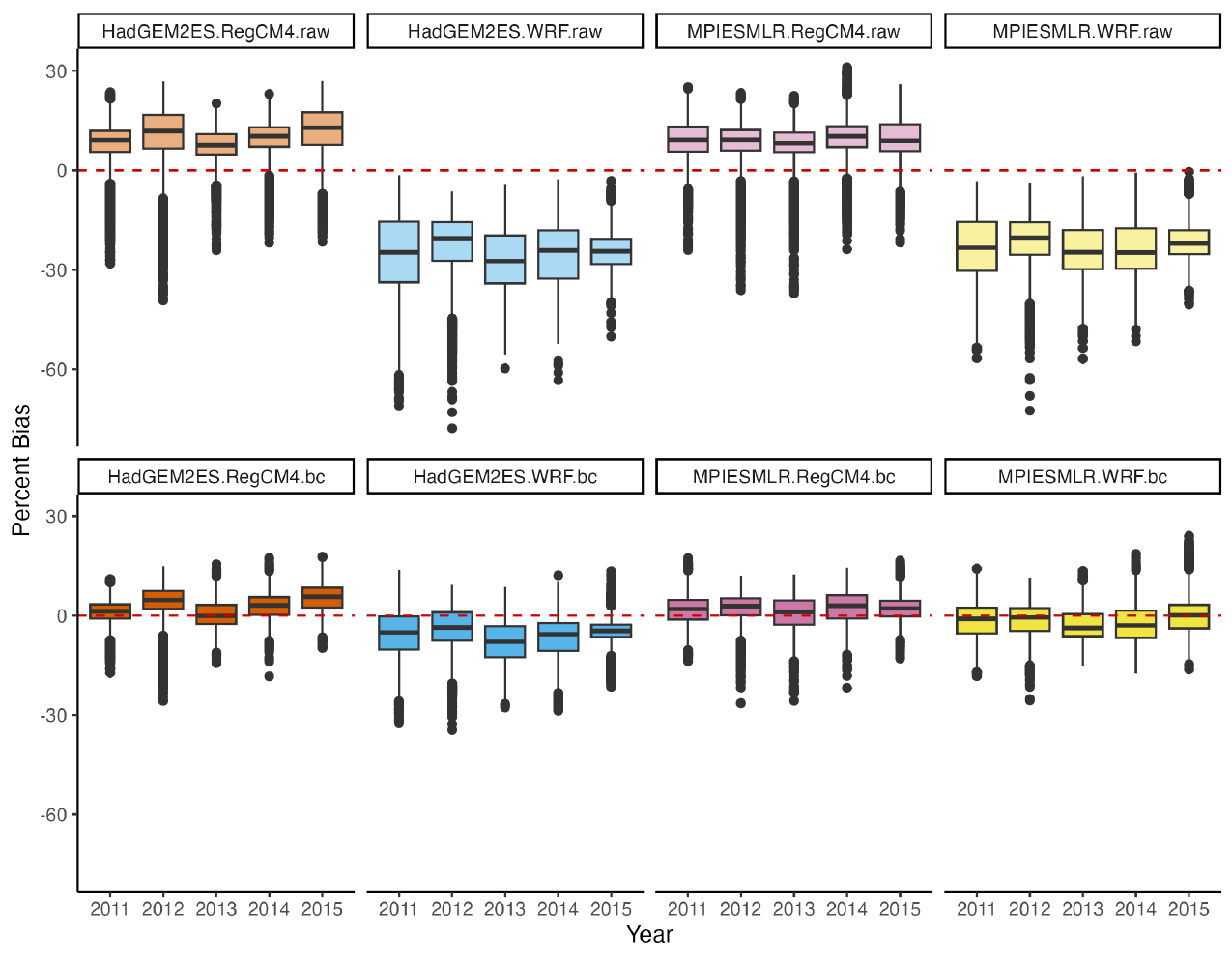}
    \includegraphics[width=0.8\textwidth]{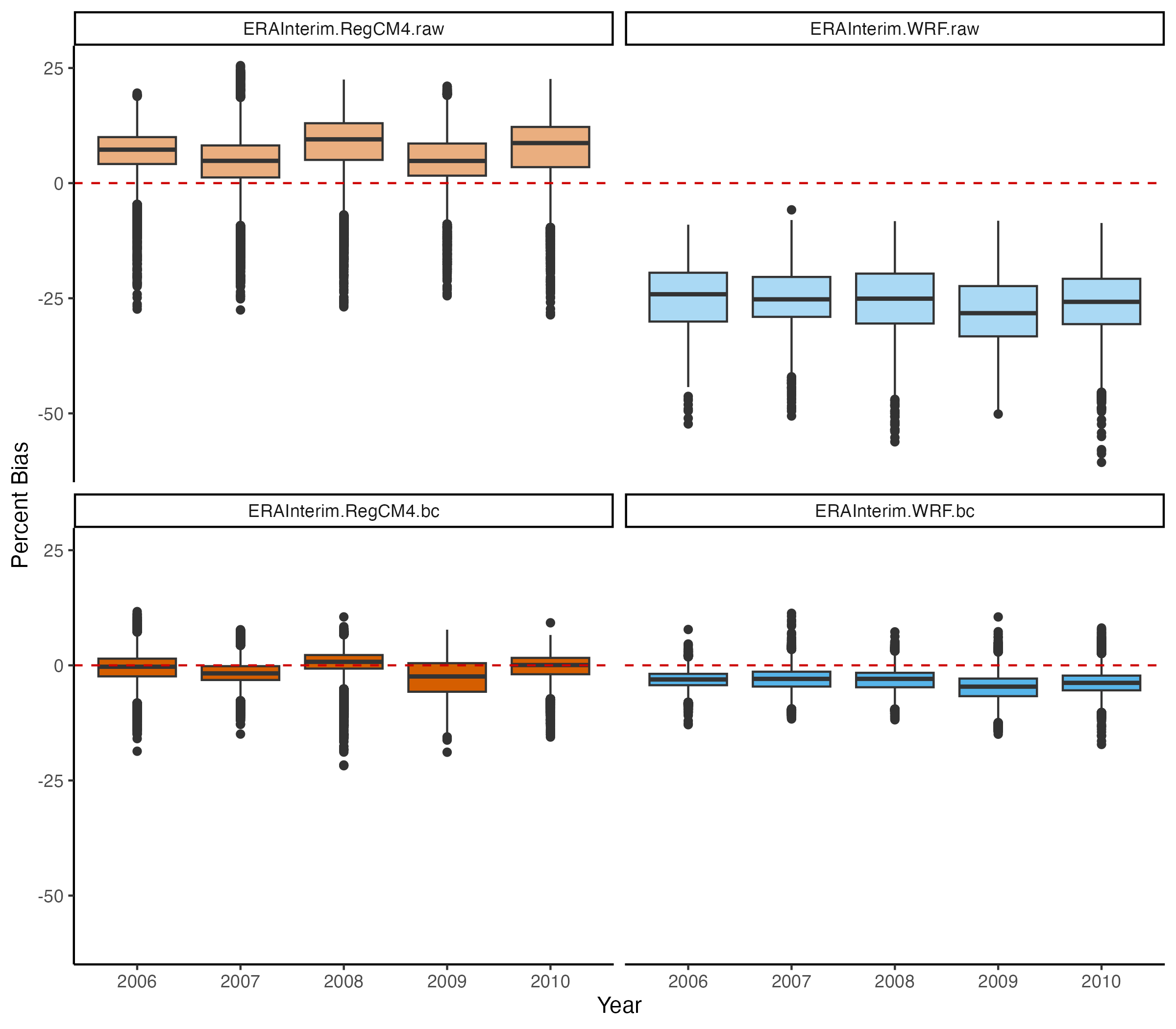}
    \caption{Percent bias of annual average $GHI (kWh/m^2/day)$ by model and year across all locations. The top row in both figures shows biases for the raw data and bottom row remaining biases for bias corrected data.}
    \label{fig:ann_avg_bias_boxplots}
\end{figure}

\begin{figure}
    \centering
    \includegraphics[width=0.75\textwidth]{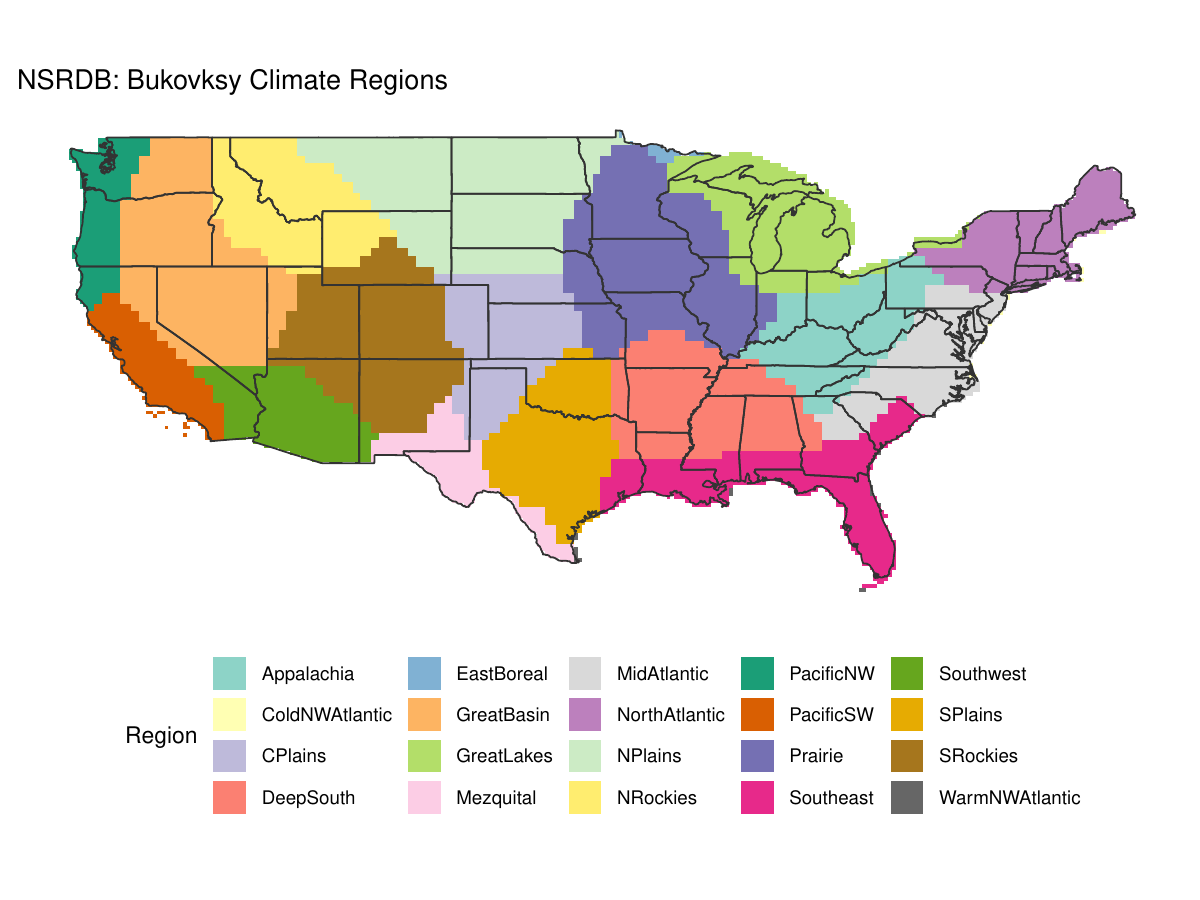}
    \caption{Defined Bukovksy regions on the NSRDB grid for CONUS.}
    \label{fig:bukovsky_regions_conus}
\end{figure}

Yearly trends in bias across all locations are plotted in  Figure~\ref{fig:ann_avg_bias_boxplots}. Again, the boxplots show that the RCM tends to control whether the model is initially over- or underpredicting GHI, matching previous results. For example, it is clear that the WRF RCM tends to over predict annual average GHI across all years compared to the RegCM4. After bias correcting through quantile mapping, most biases become centered around zero, seen in the second row. However, this result is not consistent with correction for ERA-Interim only forced runs, seen in the lower panels of Figure~\ref{fig:ann_avg_bias_boxplots}. Quantile mapping corrects well for RegCM4 however there are is still a slight overestimation of annual average GHI remaining in corrected WRF data. This follows earlier results. 

These results could indicate that the two GCMs considered are initially less biased than the RCMs or that the systematic biases introduced by the RCMs far outweigh any model biases introduced by the two GCMs considered in this analysis. This may not hold true across other GCMs from NA-CORDEX. It is also important to note that extreme or outlier biases from the original data, seen as points past the whiskers of each boxplot in the top row, are significantly reduced through bias correction. For example, uncorrected data from the WRF RCM overpredicted annual average GHI by over 60\% and this was reduced to less than 35\% for outliers, while a majority of the data was to corrected to within 15\% bias. Data from the RegCM4 models, in the orange and pink boxplots, also exemplify this correction. Here, the uncorrected data has a high amount of extreme overprediction of over 30\% for annual average GHI, despite most projections being under projections. The corrected data shows much shorter tails with percent bias for over prediction reduced to less than 30\%.

\subsection{Climate region results}
\label{sec:bc_results_clim_regions}

Analyzing biases by climate region supports earlier findings in that bias corrected WRF RCM still tends to overestimate across regions while the RegCM4 more closely matches observed patterns, seen in Figure~\ref{fig:ann_avg_bias_bukovsky_gcms}. There isn't a strong pattern seen in any one region across models, with most bias corrected data remaining within $\pm 10\%$ bias. An exception to this is the Pacific Northwest which shows higher variability in percent bias, extending to nearly $\pm 20\%$ bias for all models.  The Pacific Southwest, an area generally covering California, also sees slight underprediction of annual average GHI across all bias corrected data. However in these regions, the HadGEM-ES2.WRF has little remaining bias. 

While the results here suggest that, in general, no one region is heavily biased by any of the four models considered in this analysis, the methods developed could be useful in future analyses of solar radiation biases present in climate model data across CONUS. This sort of result could be helpful in aiding policymakers and PV system planners on determining which model may be most useful in planning. It could also point to using one model for planning in one region and another in a separate region. Finally, such analyses could aid climate model developers in identifying where geographic biases persist after bias correction, and help consider how atmospheric physics are modeled in such regions.

\begin{figure}
    \includegraphics[width=\textwidth]{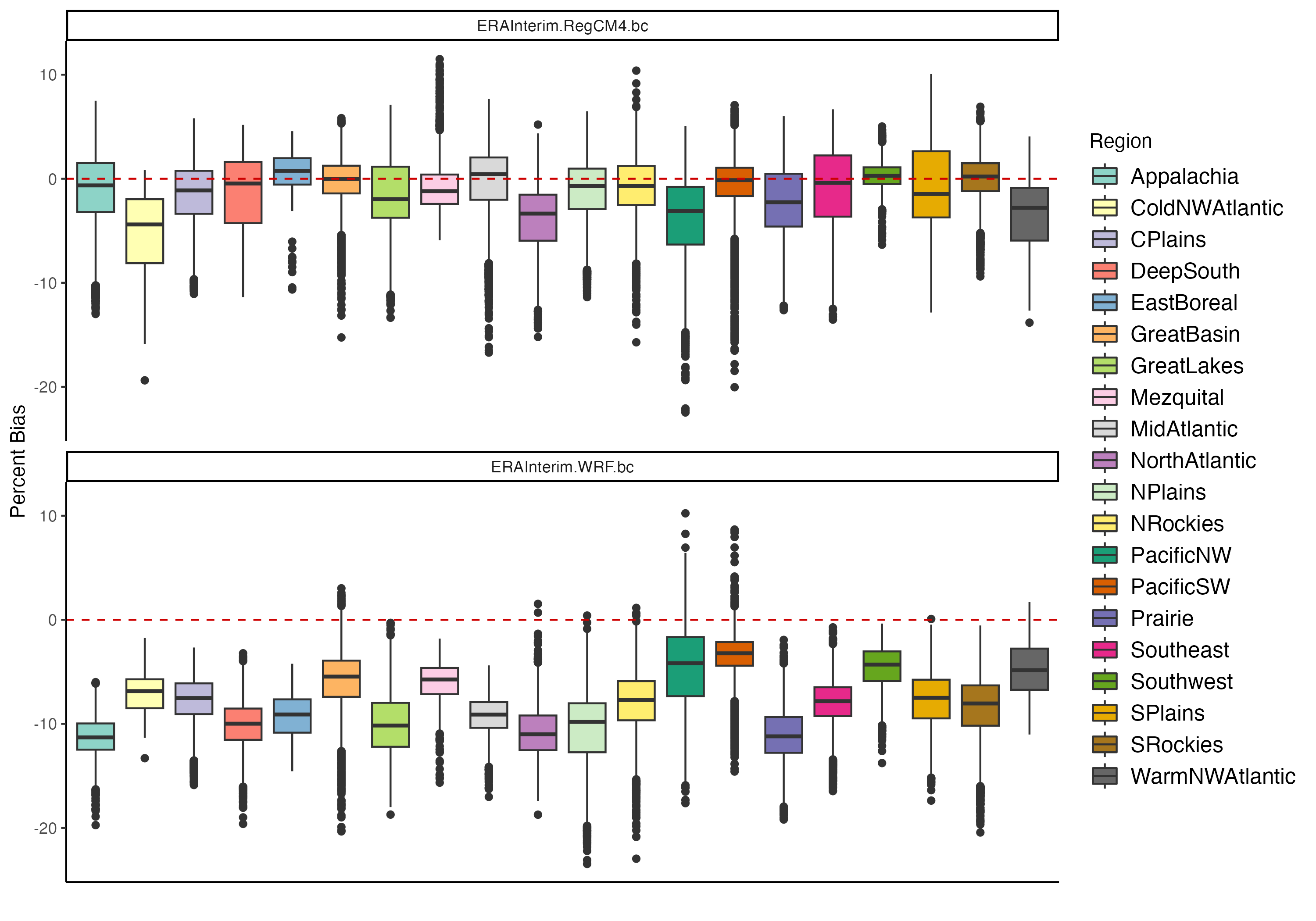}
    \caption{Percent bias after quantile mapping for climate model simulations for ERA-Interim forced WRF and RegCM4 by Bukovsky Climate Region, averaged across 2011-2015 for each region.}
    \label{fig:ann_avg_bias_bukovsky_gcms}
\end{figure}

\begin{figure}
    \centering
    \includegraphics[width=\textwidth]{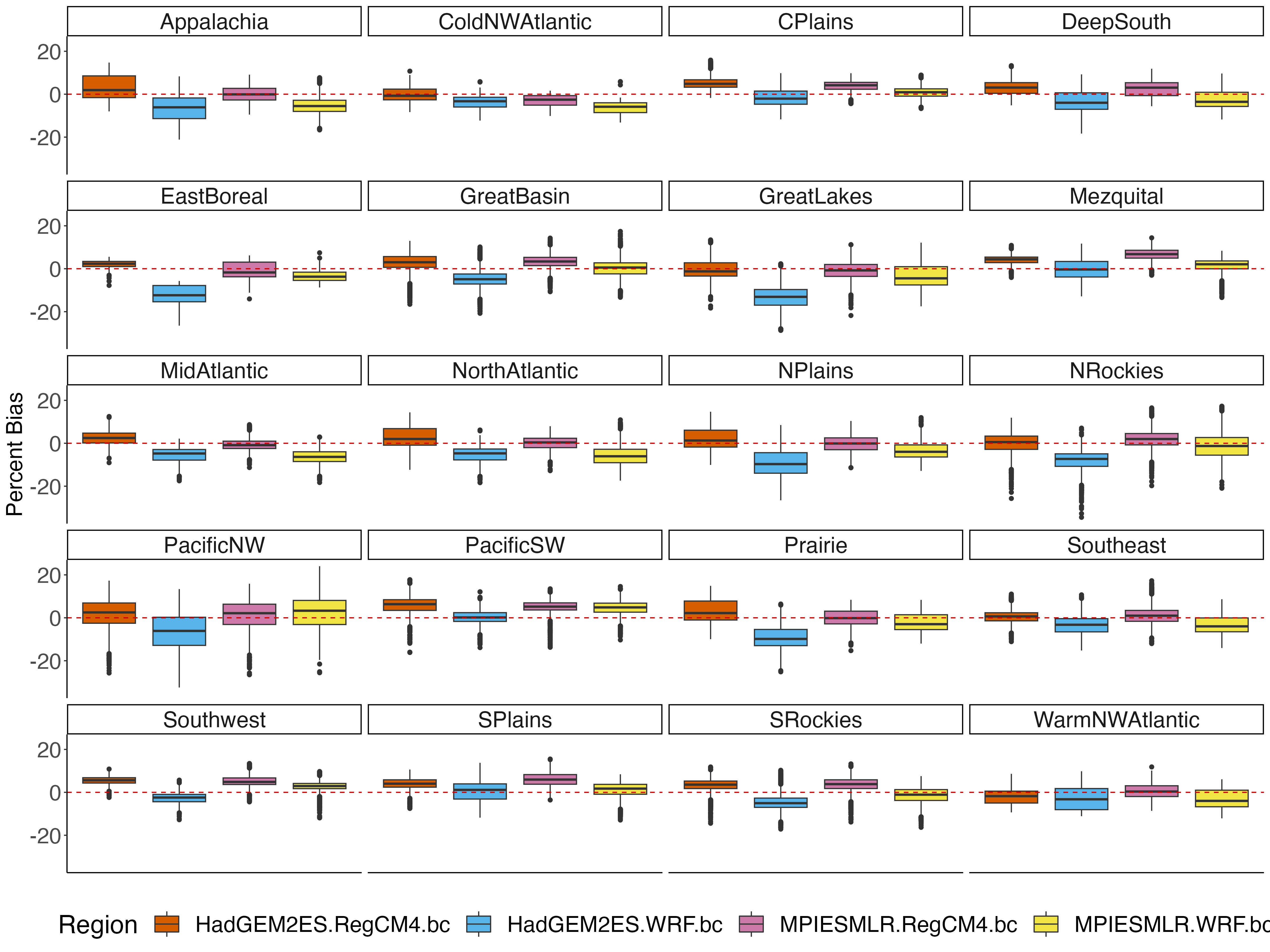}
    \caption{Bias corrected climate model simulations by Bukovsky Climate Region. Percent biases after biases correction are averaged across 2011-2015 resulting in a single bias per pixel within each region.}
    \label{fig:ann_avg_bias_bukovsky_gcms}
\end{figure}

\begin{figure}
    \centering
    \includegraphics[width = \textwidth]{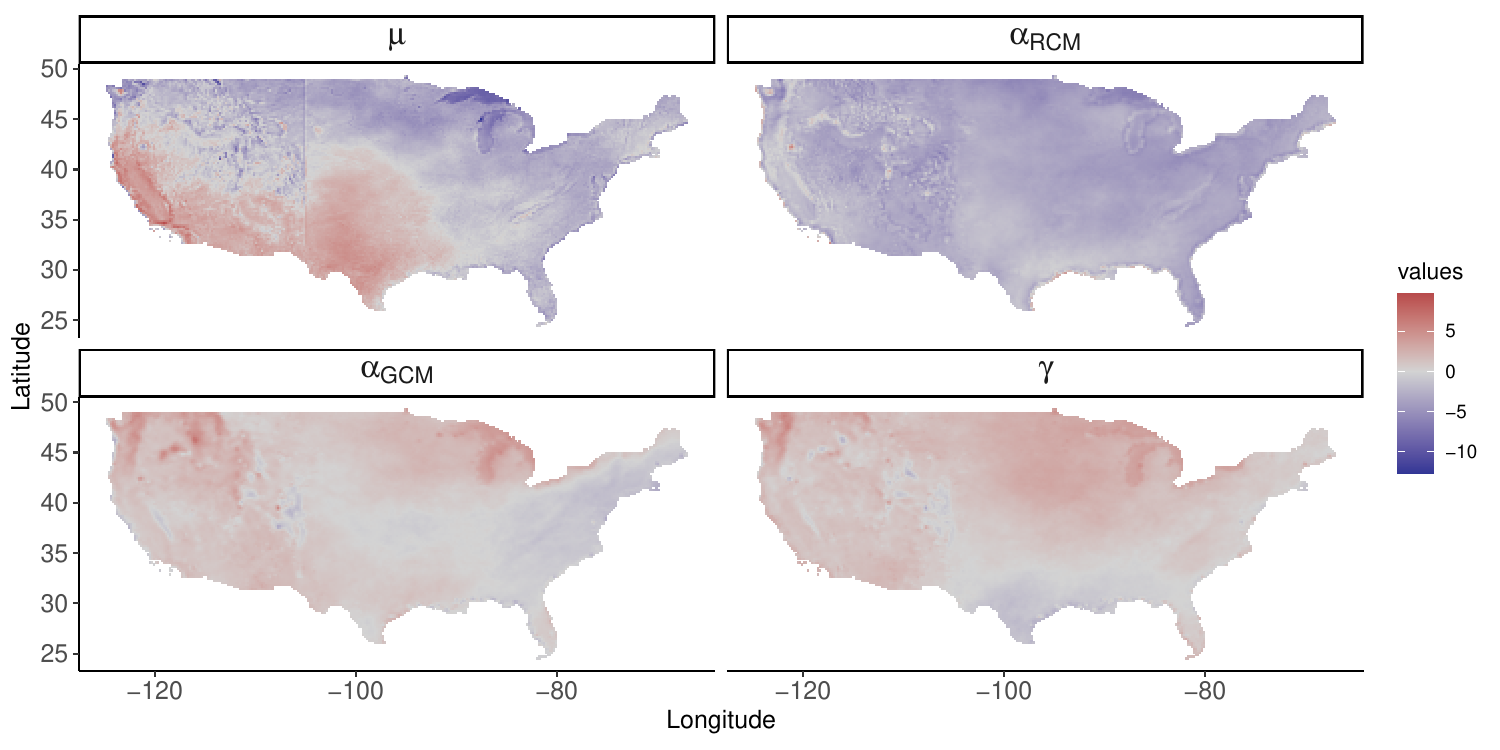}
    \includegraphics[width = \textwidth]{fanova_effects_across_2011_2015_free_scale.pdf}
    \caption{Resulting $\alpha$ components from FANOVA across all years. The top four plots are all on the same color scale while the bottom four show free scale results to show smaller-scale geographic trends.}
    \label{fig:fanova_all_years}
\end{figure}

\subsection{FANOVA}
\label{sec:bc_results_fanova}

We can solidify findings from the previous two sections through Figure~\ref{fig:fanova_all_years} which plots  FANOVA results after implementing quantile mapping. Here, a negative value indicates an overprediction of annual average solar radiation and a positive value an underprediction (bias is calculated as the bias corrected data subtracted from the observed). Note that in some maps, a vertical line appears at the -105 longitude. This results from the NSRDB combining data from two satellites at this longitude, causing the appearance of a discontinuity. The averaging has reduced the range of the biases when compared to the individual models. On average across years 2011-2015, the baseline bias ($\mu$) is centered around zero with underprediction in southeastern CONUS and overprediction in northern and eastern CONUS. This is shown in the top left plot for $\mu$ in Figure~\ref{fig:fanova_all_years}. Deviations from the overall bias that can be attributed to RCMs, GCM main effects, or the interaction of the two are shown in the top right, bottom left, and bottom right plots, respectively. 

It is clear that the RCMs tend to have a larger effect on overestimation of annual average GHI after bias correction. The top right plot in Figure~\ref{fig:fanova_all_years} shows this bias in light blue. Therefore, we could attribute remaining overestimation biases to RCMs, supporting previous findings. GCMs, on the other hand, tend to contribute little to the remaining biases and seem to contribute underestimation in small subsets of areas in CONUS, primarily the northwest and around the Great Lakes. However, the effects of bias from RCMs and GCMs seem to counteract one another in these smaller regions. 

The final component, $\gamma$, the interaction term, has some effect in northern CONUS but overall contributes less to remaining biases for 2011-2015. This suggests that bias does not depend on whether a RCM is forced by a particular GCM. Rather, we see that remaining biases are influenced by which RCM is being forced, not by which RCM is forced by which GCM.

\subsection{Clearsky Index}
\label{sec:bc_results_cs_index}

\begin{figure}
    \centering
    \includegraphics[width=\textwidth]{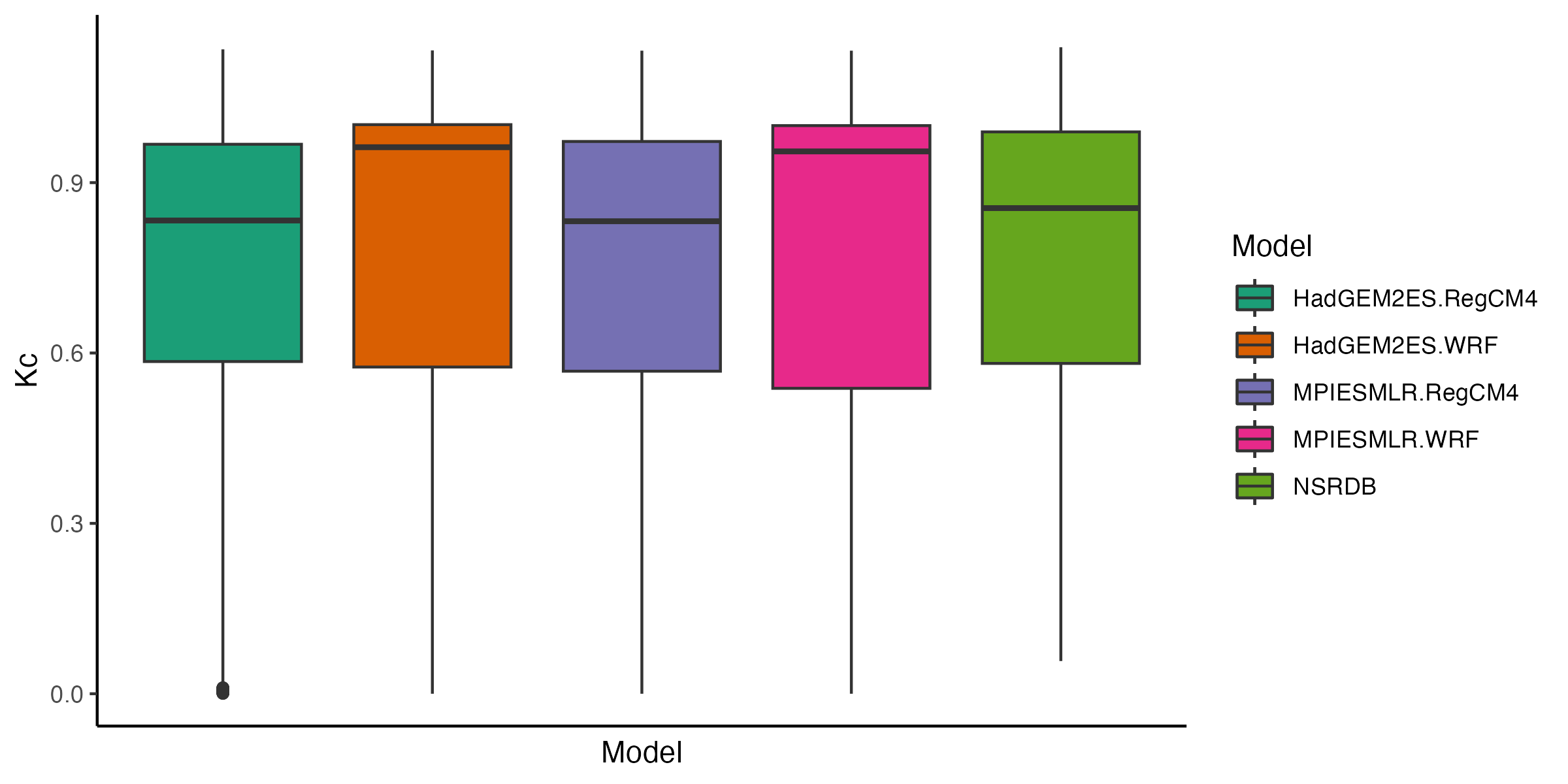}
    \caption{Bias corrected daily average clearsky index for 2011 across CONUS compared against the NSRDB.}
    \label{fig:cs_index_2011}
\end{figure}

Resulting clearsky index values for bias corrected models in 2011 are shown in Figure~\ref{fig:cs_index_2011}. GCMs that force the RegCM4 more closely match the NSRDB while the GCMs that force WRF still retain a sunny bias in the median value. The inner-quartile range of all bias corrected clearsky indices closely match the observe data. It promising that the upper limit of each boxplot is close to the same as the limit from the NSRDB, however we see that the lower limit of the bias corrected $k_c$ values extend closer to zero than the observed. A focus of this analysis was to limit GHI to physically realistic values similar to those given by observed data and these results show that this method does well to retain upper limits of GHI within physically plausible ranges but could improve on the lower limits of daily averages.

\section{Conclusion and Future Work}
\label{sec:bc_conclusions}

This chapter introduces a novel adaptation of quantile mapping specifically for GHI. The method aims to control the upper limit of GHI by transforming data into a clearsky index and then again using the logit transformation. Using a higher number of equally spaced quantiles in $[0,1]$ corrected well overall for monthly biases as well as annual average GHI. The FANOVA approach reveals the source of persistent biases, particularly underestimation, result from RCMs which have stronger effects than GCMs or the interaction of both. We saw that geographic biases that remain from quantile mapping may be artifacts of certain years or a particular RCM rather than resulting from the GCM or specific combination of GCMs and RCMs. Overall, RCMs seem to contribute most to biases seen in the chosen model data forced by GCMs and are generally corrected for after bias correction. This suggests that while RCMs are a valuable dynamical downscaling method, they may introduce strong systematic biases. Data from RCMs used for planning solar facilities, particularly under various climate path scenarios, should first be bias corrected. This is particularly true in the application of analyzing GHI data, where there is an upper limit for physically reasonable values. 

While this method is a promising adaptation of quantile mapping, there could be some improvements. Using clearsky data to enforce reasonable upper limits on the bias corrected data seemed to work well, but bias correction for data extending into future years relies on an empirically derived clearsky dataset. There is no guarantee that the dataset used in this study will represent clearsky data several decades from now. Therefore, the development of a clearsky dataset for future years may be worthwhile, if needed for other applications in addition to this study. Second, this study focused on using a 2$\times$2 factorial design in order to be able to compare biases between GCMs and RCMs and implement a FANOVA methodology. Although these results are specific to these models, the methodology could be easily adaptable to other RCMs, where available. Additionally, we saw that there weren't strong biases for any particular climate region when using Bukovsky regions. Therefore, future work could examine how well the quantile mapping method performs conditional on GHI values rather than climate.

\appendix
\section{Additional Figures}
\label{appendix:bc_additional_figures}

\begin{figure}[h]
    \centering
    \includegraphics[width=\textwidth]{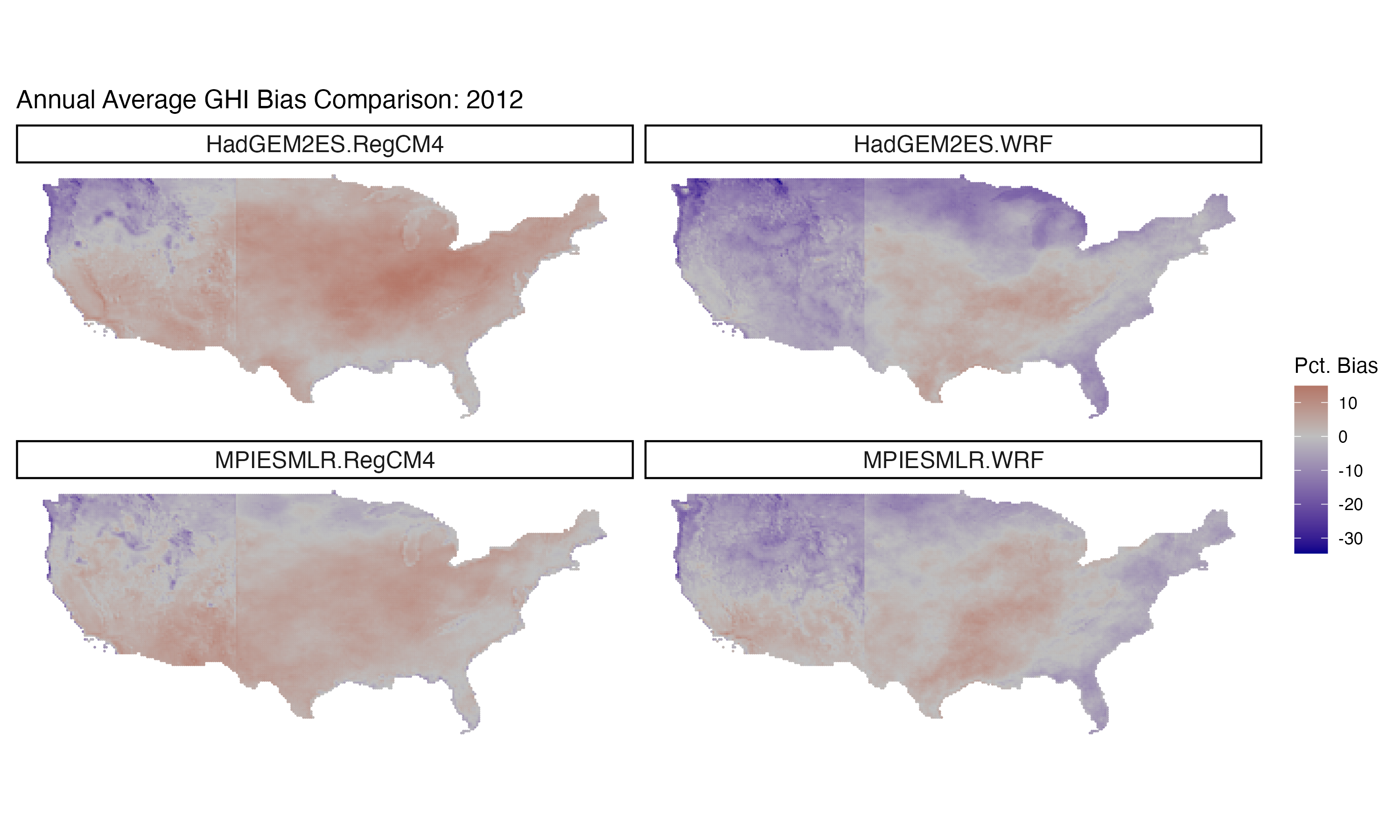}
    \caption{Percent bias in annual average GHI between bias corrected climate model output from NA-CORDEX and observed data from the NSRDB for entire CONUS in 2012. Similar trends appear depending on the RCM, where we see an underprediction or GHI in southern CONUS and midwest CONUS and overpredicted of annual average GHI in northwestern CONUS for WRF. This trend is similar for the RegCM4 RCM but the underprediction of annual average GHI is more widespread.}
\end{figure}

\begin{figure}[h]
    \centering
    \includegraphics[width=\textwidth]{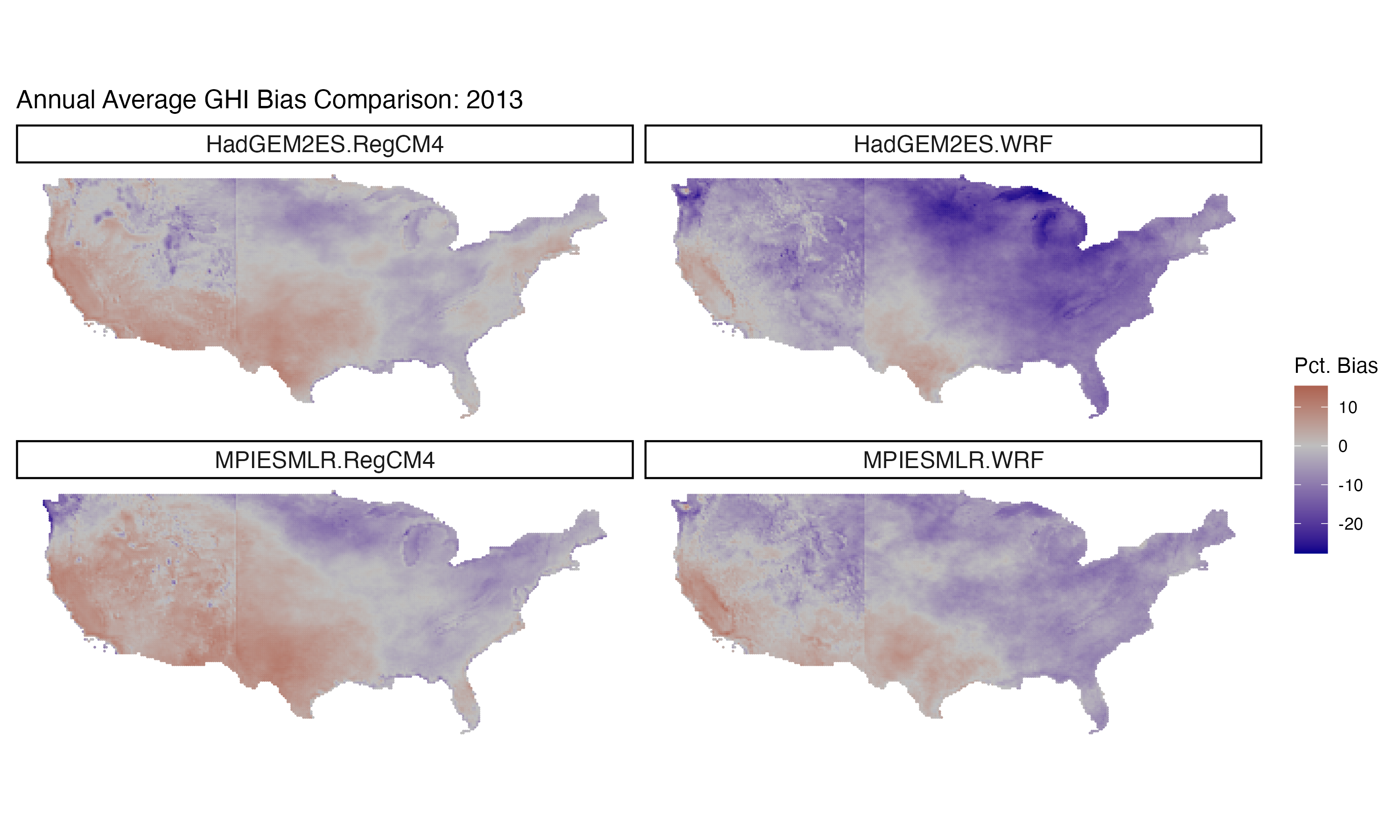}    \caption{Percent bias in annual average GHI between bias corrected climate model output from NA-CORDEX and observed data from the NSRDB for entire CONUS in 2013. Here, underprediction for the WRF RCM is concentrated to the southern Texas and California while overprediction for the RegCM4 covers most of southwestern and western CONUS. For this particular year, underprediction appears to occur in the northeast and eastern CONUS for models with the WRF RCM.}
\end{figure}

\begin{figure}[h]
    \centering
 \includegraphics[width=\textwidth]{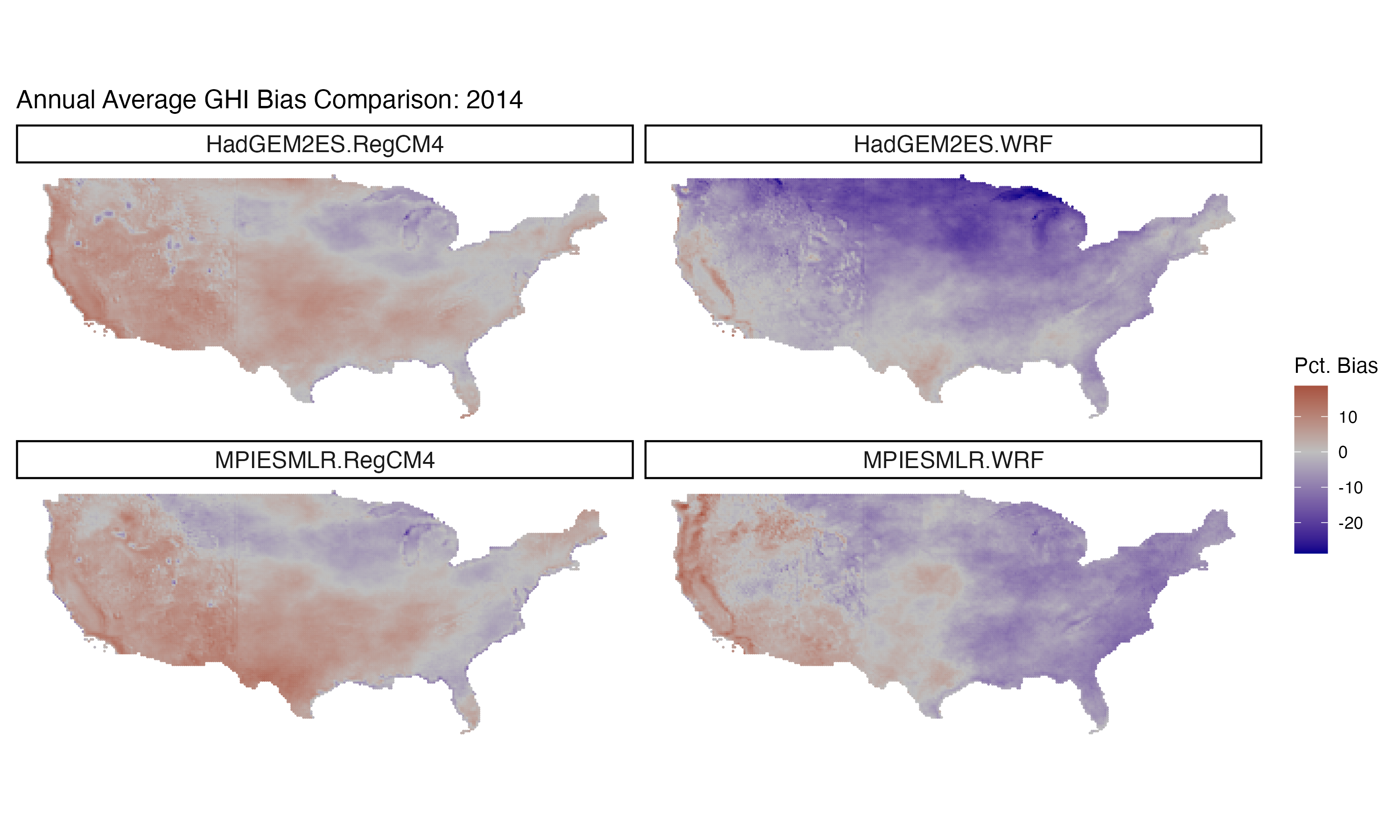}    \caption{Percent bias in annual average GHI between bias corrected climate model output from NA-CORDEX and observed data from the NSRDB for entire CONUS in 2014. Underprediction of annual average GHI again occurs in southern and western CONUS for this year. For the WRF RCM, overprediction tends to occur in the northeast and eastern areas of CONUS.}
\end{figure}

\begin{figure}[h]
    \centering
 \includegraphics[width=\textwidth]{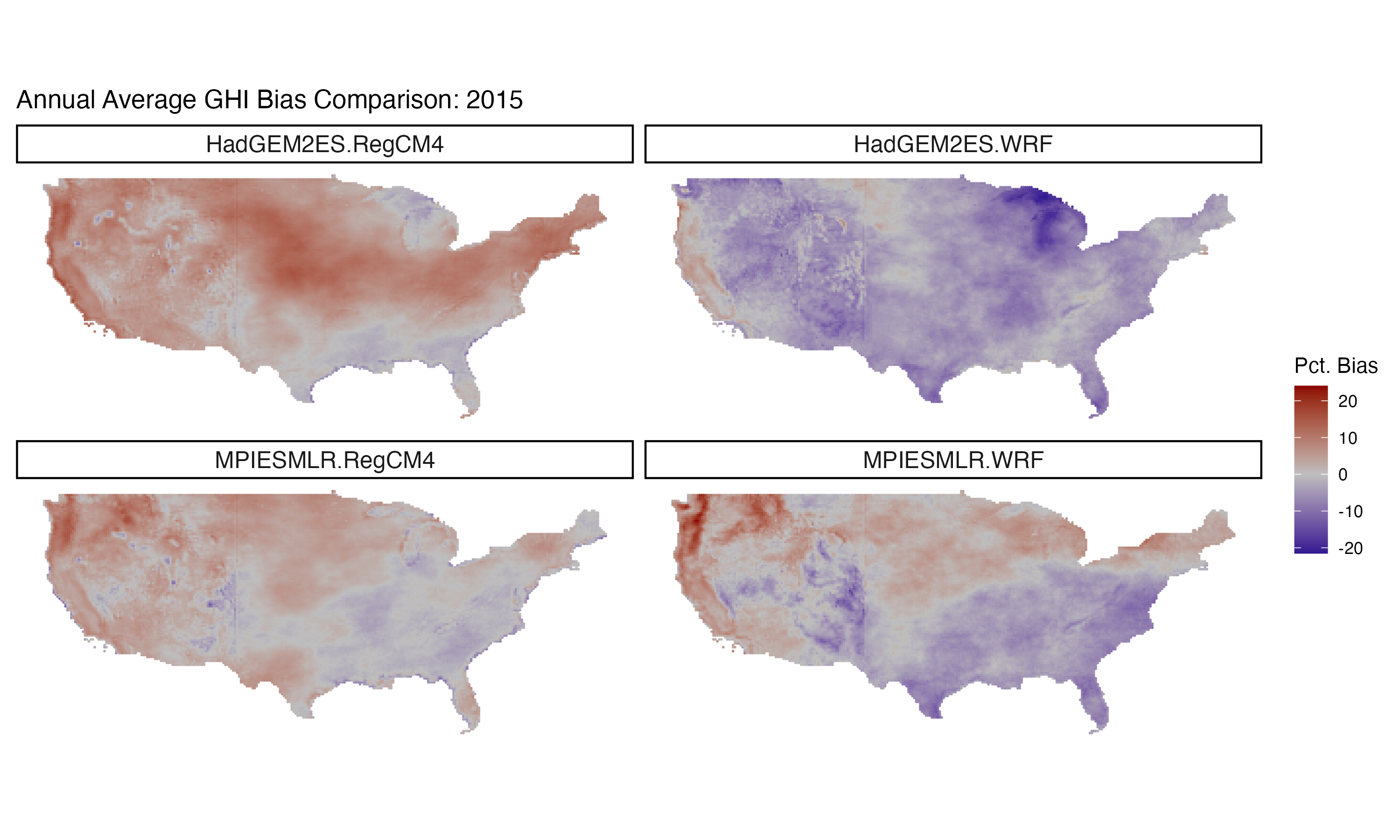}    \caption{Percent bias in annual average GHI between bias corrected climate model output from NA-CORDEX and observed data from the NSRDB for entire CONUS in 2015. Patterns from previous years appear flipped for the WRF RCM, where now there is an overprediction of annual average GHI in southern CONUS and underprediction in northern and northwest CONUS. For this particular year, the RegCM4 widely underpredicts annual average GHI.}
\end{figure}

\begin{figure}[h]
    \centering
    \includegraphics[width=\textwidth]{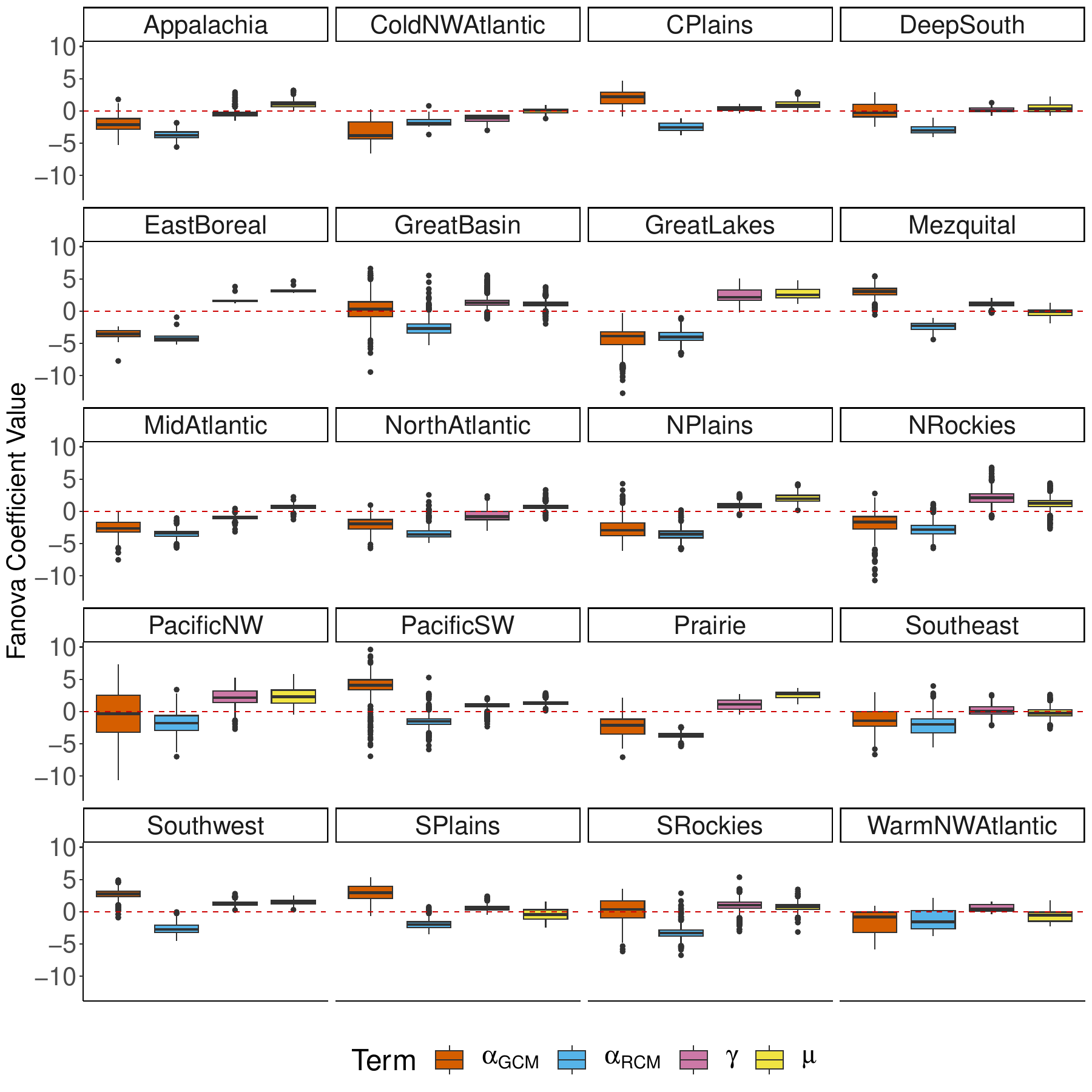}
    \caption{Fanova coefficient value by Bukovsky climate region.}
    \label{fig:fanova_all_years_by_clim_region}
\end{figure}





\clearpage

\bibliographystyle{elsarticle-harv} 
\bibliography{qm_solar}







\end{document}